\documentclass[prd,showpacs,preprintnumbers,nofootinbib,aps]{revtex4}
\usepackage{subeqnarray}
\usepackage{epsfig,amsmath,amssymb}
\usepackage{mathrsfs}
\usepackage[usenames,dvipsnames]{color}
\usepackage[pagebackref=false, colorlinks=false]{hyperref}
\definecolor{redish}{rgb}{0.7,0.2,0.0}  
\definecolor{bluish}{rgb}{0.2,0.5,0.8}
\hypersetup{linkcolor=redish,          
                  citecolor=blue,        
                  filecolor=magenta,      
                  urlcolor=bluish}          
\DeclareFontFamily{U}{rsfs}{}         
\DeclareFontShape{U}{rsfs}{m}{n}{<5> rsfs5 <6><7> rsfs7          %
  <8><9><10><10.95><12><14.4><17.28><20.74><24.88> rsfs10}{}     %
\DeclareMathAlphabet{\mathfs}{U}{rsfs}{m}{n}                     %

\newcommand{\ba}{\nopagebreak[3]\begin{eqnarray}}
\newcommand{\ea}{\end{eqnarray}}
\newcommand{\bii}{\begin{itemize}}
\newcommand{\eii}{\end{itemize}}

\begin{document}

\title{Thermal Fluctuations Of Stable Quantum ADS Kerr-Newman Black Hole}
\author{Aloke Kumar Sinha}
\email{akshooghly@gmail.com}
\affiliation{ Haldia Government College,West Bengal, India}
       
\begin{abstract}

 We have already derived the Criteria for thermal stability of charged rotating black holes in any dimension , for horizon areas that are large relative to the Planck area (in these dimensions). The derivation is done by using results of loop quantum gravity and equilibrium statistical mechanics of  the Grand Canonical ensemble. It is also shown there \cite{aspm 16} that in four dimensional spacetime, ADS Kerr-Newman Black hole is thermally stable. In this paper, the expectation values of fluctuations of horizon area,charge and angular momentum of stable ADS black hole are calculated. Interestingly, it is found that leading order fluctuations of charge and angular momentum , in large horizon area limit ,  are independent of the values of charge and angular momentum at equilibrium.

\end{abstract}
\maketitle

\section{Introduction}

 Semiclassical analysis informs that nonextremal, asymptotically flat black holes are thermally unstable  due to decay under Hawking radiation, with negative specific heat  \cite{dav77}. This motivated the study of thermal stability of black holes, from a perspective that relies on a definite proposal for {\it quantum} spacetime (like Loop Quantum Gravity, \cite{rov,thie}) . A consistent understanding  of {\it quantum} black hole entropy has been obtained through Loop Quantum Gravity \cite{abck98,abk00}, where not only has the Bekenstein-Hawking area law been retrieved for macroscopic (astrophysical) black holes, but a whole slew of corrections to it, due to quantum spacetime fluctuations have been derived as well \cite{km98}-\cite{bkm10}, with the leading correction being logarithmic in area with the coefficient $-3/2$.\\
 
 Classically a black hole in general relativity is characterized by its' mass ($M$), charge ($Q$) and angular momentum ($J$).  Intuitively, therefore, we expect that thermal behaviour of black holes will depend on all of these parameters. The simplest case of vanishing charge and angular momentum has been investigated longer than a decade ago \cite{cm04} - \cite{cm05-2} and that has been  generalized, via the idea of {\it thermal} holography \cite{pm07}, \cite{pm09}, and the saddle point approximation to evaluate the canonical partition function corresponding to the horizon, retaining Gaussian thermal fluctuations.  This body of work has been generalized first \cite{mm12} to include black holes with charge and recently \cite{aspm 16} for charged rotating black holes. There it is shown that  anti-de Sitter(ADS) Kerr-Newman black hole (for a range of cosmological constants) is thermally stable . \\

In this paper, using previous knowledge, expectation values of thermal fluctuations of all the hairs i.e. charge, horizon area and mass are calculated.These are calculated in the limit of large horizon area. \\  

The paper is organized as follows: In section 2, the idea of thermal holography, alongwith the concept of (holographic) mass associated with horizon of a black hole is briefly reviewed and also the revision of grand canonical partition function of charged rotating black hole (ADS Kerr-Newman Black Hole) and condition for its' thermal stability are done .  In the next section, detailed calculation of thermal fluctuations are done for ADS Kerr-Newman black hole. Last section contains a brief summary and outlook.

\vspace{.3 in}
\section{Thermal holography}

 In this section, we briefly review the necessary part of our earlier work for present purpose and hence some overlapping with \cite{aspm 16} is inevitable. 

\subsection{Mass Associated With horizon}

Black holes at equilibrium are represented by isolated horizons, which are internal boundaries of spacetime. Hamiltonian evolution of this spacetime gives the first law associated with isolated horizon($b$) and is given as,
\begin{eqnarray}
\delta E^{t}_{h}=\frac{\kappa^{t}}{8\pi}\delta A_{h}+\Phi^{t}\delta Q_{h}+\Omega^{t}\delta J_{h}
\end{eqnarray}
where, $E^{t}_{h}$ is the energy function associated with  the horizon, $\kappa^{t}$, $\Phi^{t}$ and $\Omega^{t}$ are respectively the surface gravity, electric potential and   angular velocity of  the horizon; $Q_{h}~,~A_{h}$ and $J_{h}$ are respectively the charge, area and angular momentum of  the horizon. The label '$t$' denotes the particular time evolution field ($t^{\mu}$)  associated with the spatial hypersurface chosen. This hypersurface foliates the horizon. $E^{t}_{h}$ is  assumed here to be a function of $A_{h}$, $Q_{h}$ and $J_{h}$.

 As argued in \cite{aspm 16} , mass can be defined on the isolated horizon.

\subsection{Quantum Geometry}

The Hilbert space of  a generic quantum spacetime is given as, $\mathcal{H}=\mathcal{H}_{b}{\otimes}\mathcal{H}_{v}$ , where $b(v)$ denotes the boundary (bulk)  space. A generic quantum state is  thus given as
\begin{equation}
\vert\Psi\rangle=\sum\limits_{b,v} C_{b,v} \vert\chi_{b}\rangle {\otimes} \vert\psi_{v}\rangle ~\label{genstate} 
\end{equation} 
Now, the full Hamiltonian operator ($\widehat{H}$),  operating on $\mathcal{H}$ is given by
\begin{equation}\label{hamil}
\widehat{H}\vert\Psi\rangle=(\widehat{H_{b}}{\otimes}I_{v}+I_{b}{\otimes}\widehat{H_{v}})\vert\Psi\rangle
\end{equation} 
where,  respectively, $I_{b} (I_{v})$ are identity operators on $\mathcal{H}_{b} (\mathcal{H}_{v})$ and $\widehat{H_{b}} (\widehat{H_{v}})$ are the Hamiltonian operators on $\mathcal{H}_{b}(\mathcal{H}_{v})$. 

 So, in this case the first class constraints are realized as 
\begin{eqnarray}
[\widehat{H_{v}} - \Phi \widehat{Q_{v}} - \Omega \widehat{J_{v}}] | \psi_{v} \rangle = 0 ~.~\label{fullham}
\end{eqnarray}

\subsection{Grand Canonical Partition Function}

Consider the black hole immersed in a heat bath, at some (inverse) temperature $\beta$, with which it can exchange energy, charge and angular momentum. The grand canonical partition function of the black hole is given as,

\begin{eqnarray}
Z_{G}=Tr(exp(-\beta\widehat{H}+\beta\Phi\widehat{Q}+\beta\Omega\widehat{J})) ~\label{gcpf}
\end{eqnarray}
where  the trace is taken over all states. \\

The application of the Poisson resummation formula \cite{cm04} changes the form this grand canonical partition function as\cite{aspm 16},
\begin{equation}
Z_{G}=\int dA\hspace{.05 in} dQ\hspace{.05 in} dJ\hspace{.05 in} \exp [S(A)-\beta(E(A,Q,J)-\Phi Q-\Omega J)]~ \label{pfresult}
\end{equation}
where, $S(A)$ is the microcanonical entropy of the horizon.

\subsection{Saddle Point Approximation and Stability Criteria}

 The equilibrium configuration of black hole is given by the saddle point $\bar{A},\bar{Q},\bar{J}$ in the three dimensional space of integration over area, charge and angular momentum. This configuration is identified with with an {\it isolated} horizon with fluctuations $a=(A-\bar{A}), q=(Q-\bar{Q}),j=(J-\bar{J})$ around the saddle point.  Taylor expanding eqn (\ref{pfresult}) about the saddle point, yields 
\begin{eqnarray}
Z_{G} &=& \exp[ S(\bar{A})-\beta M(\bar{A},\bar{Q},\bar{J})+\beta\Phi \bar{Q}+\beta\Omega\bar{J}] \nonumber \\
&\times & \int da~ dq~ dj~ \exp \{-\frac{\beta}{2}[( M_{AA}-\frac{S_{AA}}{\beta} )a^{2} + ( M_{QQ})q^{2}+(2 M_{AQ})aq \nonumber \\
&+& ( M_{JJ})j^{2}+(2M_{AJ})aj+(2 M_{QJ})qj] \} ~ \label{sadpt}
\end{eqnarray}
where  meaning of all the terms are as given in \cite{aspm 16}. \\

Convergence of the integral (\ref{sadpt}) implies  that the Hessian matrix ($H$) has to be positive definite, where
\begin{eqnarray}
 H = \left( \begin{array}{ccc}
\beta M_{AA}(\bar{A},\bar{Q},\bar{J})- S_{AA}(\bar{A},\bar{Q},\bar{J}) \hspace{.3 in}& \beta M_{AQ}(\bar{A},\bar{Q},\bar{J}) \hspace{.3 in} & \beta M_{AJ}(\bar{A},\bar{Q},\bar{J})\vspace{.1 in} \\  
\beta M_{AQ}(\bar{A},\bar{Q},\bar{J}) \hspace{.3 in}& \beta M_{QQ}(\bar{A},\bar{Q},\bar{J}) \hspace{.3 in}& \beta M_{JQ}(\bar{A},\bar{Q},\bar{J}) \vspace{.1 in} \\
\beta M_{AJ}(\bar{A},\bar{Q},\bar{J}) \hspace{.3 in}& \beta M_{JQ}(\bar{A},\bar{Q},\bar{J}) \hspace{.3 in}& \beta M_{JJ}(\bar{A},\bar{Q},\bar{J}) \end{array} \right) \label{hess}
\end{eqnarray}
The necessary and sufficient conditions for  a real symmetric square matrix to be positive definite is : 'determinants all principal square submatrices, and the determinant of the full matrix, are positive.'\cite{meyer} This condition leads to the  `stability criteria'  that are described in \cite{aspm 16}.Ofcourse, (inverse) temperature $\beta$ is assumed to be positive for a stable configuration.   The microcanonical entropy for  {\it macroscopic} isolated horizons($S$) has the form
\begin{eqnarray}
S~&=&~S_{BH} ~-~\frac32 \log S_{BH} +{\cal O}(S_{BH}^{-1}) ~\label{kment} \\
S_{BH} ~&=& ~ \frac{A_h}{4 A_P}~,~A_P \rightarrow {\rm Planck~area} ~. \label{bek}
\end{eqnarray}

\section{Thermal Fluctuations Of ADS Kerr-Newman Black Hole}

 The expectation value of fluctuation of any quantitity is the standard deviation of that quantity. It is a statistical measure of deviation of any distribution. The knowledge of probability theory and the last expression of grand canonical partition function (\ref{sadpt}) together give the standard deviation of charge($Q$) as,
 \begin{eqnarray}
(\Delta Q)^{2} = \frac{\int da~ dq~ dj~  q^{2}\exp \{-\frac{\beta}{2}[( M_{AA}-\frac{S_{AA}}{\beta} )a^{2} + ( M_{QQ})q^{2}+(2 M_{AQ})aq +( M_{JJ})j^{2}+(2M_{AJ})aj+(2 M_{QJ})qj] \}}{\int da~ dq~ dj~   \exp \{-\frac{\beta}{2}[( M_{AA}-\frac{S_{AA}}{\beta} )a^{2} + ( M_{QQ})q^{2}+(2 M_{AQ})aq + ( M_{JJ})j^{2}+(2M_{AJ})aj+(2 M_{QJ})qj] \}} ~ ~ ~ \label{chflu}
\end{eqnarray}
where, $\Delta Q$ is the standard deviation of black hole charge. Similarly, $\Delta A$ and $\Delta J$ are defined for horizon area and angular mementum of the black hole.\\

The expression(\ref{sadpt}) and (\ref{chflu}) together give,
\begin{eqnarray}
(\Delta Q)^{2} &=& -\frac{2}{\beta}\cdot\frac{1}{ Z_{G}}\cdot\frac{\partial Z_{G}}{\partial M_{QQ}} \nonumber\\
&=& -\frac{2}{\beta}\cdot\frac{\partial log Z_{G}}{\partial M_{QQ}} \label{chflu1}
\end{eqnarray}
Similarly, $(\Delta A)^{2}$ and $(\Delta J)^{2}$ are defined by taking partial derivatives with respect to $ (M_{AA}-\frac{S_{AA}}{\beta}) $ and $M_{JJ}$ respectively i.e.
\begin{eqnarray}
(\Delta A)^{2} &=& -\frac{2}{\beta}\cdot\frac{1}{ Z_{G}}\cdot\frac{\partial Z_{G}}{\partial (M_{AA}- \frac{S_{AA}}{\beta})} \nonumber\\
&=& -\frac{2}{\beta}\cdot\frac{\partial log Z_{G}}{\partial (M_{AA}- \frac{S_{AA}}{\beta})} \label{arflu}
\end{eqnarray}
\begin{eqnarray}
(\Delta J)^{2} &=& -\frac{2}{\beta}\cdot\frac{1}{ Z_{G}}\cdot\frac{\partial Z_{G}}{\partial M_{JJ}} \nonumber\\
&=& -\frac{2}{\beta}\cdot\frac{\partial log Z_{G}}{\partial M_{JJ}} \label{amflu}
\end{eqnarray}
Equation no. (\ref{sadpt}), (\ref{hess}) and (\ref{chflu1}) together give,
\begin{eqnarray}
(\Delta Q)^{2} = \frac{(\beta M_{AA}- S_{AA})\cdot \beta M_{JJ} - (\beta M_{AJ})^{2}}{\vert H \vert} \label{chflu2}
\end{eqnarray}
where, $\vert H \vert$ is the determinant of the hessian matrix($H$). \\

Equation no. (\ref{sadpt}), (\ref{hess}) and (\ref{arflu}) together give,
\begin{eqnarray}
(\Delta A)^{2} = \frac{\beta ^{2} \big(M_{QQ}M_{JJ} - (M_{JQ})^{2}\big)}{\vert H \vert} \label{arflu1}
\end{eqnarray}
Equation no. (\ref{sadpt}), (\ref{hess}) and (\ref{amflu}) together give,
\begin{eqnarray}
(\Delta J)^{2} = \frac{(\beta M_{AA}- S_{AA})\cdot \beta M_{QQ} - (\beta M_{AQ})^{2}}{\vert H \vert} \label{amflu1}
\end{eqnarray}

The AdS Kerr-Newman black hole is given in Boyer–Lindquist coordinates as 
\begin{equation}\label{adsknmetric}
ds^{2}= -\frac{\Delta_{r}}{\rho^{2}}(dt-\frac{a\hspace{.02 in} sin^{2}\theta}{\Sigma}\hspace{.02 in} d\phi)^{2} +\frac{\Delta_{\theta}\hspace{.02 in} sin^{2}\theta}{\rho^{2}}(\frac{r^{2}+a^{2}}{\Sigma} d\phi -a\vspace{.02 in}dt)^{2} +\frac{\rho^{2}}{\Delta_{r}}dr^{2}+ \frac{\rho^{2}}{\Delta_{\theta}} d\theta ^{2}
\end{equation}
where, $ \Sigma= 1-\frac{a^{2}}{l^{2}},\hspace{.1 in} \Delta_{r}=(r^{2}+ a^{2})(1+\frac{r^{2}}{l^{2}})-2\hspace{.02 in}M \hspace{.02 in}r +Q^{2} ,\hspace{.1 in}\Delta_{\theta}= 1-\frac{a^{2}cos^{2}\theta}{l^{2}},\hspace{.1in} \rho^{2}= r^{2}+a^{2}\hspace{.02 in}cos^{2}\theta  , \hspace{.1 in}  a=\frac{J}{M} $. The generalized Smarr formula for the AdS Kerr-Newman Black Hole is given as \cite{cck} 
\begin{equation}\label{adsknmass}
M^{2}=\frac{A}{16\pi}+\frac{\pi}{A}(4J^{2}+Q^{4})+\frac{Q^{2}}{2}+\frac{J^{2}}{l^{2}}+\frac{A}{8\pi l^{2}}(Q^{2}+\frac{A}{4\pi}+\frac{A^{2}}{32\pi^{2}l^{2}})
\end{equation}
where  the cosmological constant ($\Lambda$) is defined in terms of a cosmic length parameter as $\Lambda = -1/l^2$.

As before, our interest is in astrophysical (macroscopic) charged, rotating black holes whose horizon area exceeds by far both the cosmic `area' $l^2$ and the Planck area. So, in large area limit for finite charge($Q$) and angular momentum($J$) i.e.  $A >> l^2$ and  $A^2 >>4J^2 + Q^4$, one can approximate (\ref{adsknmass}) as follows 
\begin{equation}\label{adsknmassapp}
M\thickapprox\frac{A^{3/2}}{16\pi^{3/2}l^{2}}+\frac{A^{1/2}}{4\pi^{1/2}}+\frac{\pi^{1/2}Q^{2}}{A^{1/2}}+\frac{8\pi^{3/2}J^{2}}{A^{3/2}} ~.
\end{equation}
 Equation no. (\ref{hess}), (\ref{chflu2}) and (\ref{adsknmassapp}) together give,
 \begin{eqnarray}
(\Delta Q)^{2} \approx \frac{3 A_{p}A}{16 \pi^{2}l^{2}} \label{chflu3}
\end{eqnarray}
Equation no. (\ref{hess}), (\ref{arflu1}) and (\ref{adsknmassapp}) together give,
 \begin{eqnarray}
(\Delta A)^{2} \approx 8A_{p}A \label{arflu2}
\end{eqnarray}
Equation no. (\ref{hess}), (\ref{amflu1}) and (\ref{adsknmassapp}) together give,
 \begin{eqnarray}
(\Delta J)^{2} \approx \frac{3 A_{p}A^{2}}{128 \pi^{3}l^{2}} \label{amflu2}
\end{eqnarray}
 Ofcourse, last three expressions are only the leading order terms in large horizon area limit.

\section{Summary and Discussion}
We reiterate that our analysis is quite independent of specific classical spacetime geometries, relying as it does on quantum aspects of spacetime. The construction of the partition function used standard formulations of equilibrium statistical mechanics augmented by results from canonical Quantum Gravity, with extra inputs regarding the behaviour of the microcanonical entropy as a function of area {\it beyond the Bekenstein-Hawking area law}, as for instance derived from Loop Quantum Gravity \cite{km00}. We use classical metric only as an input which gives the dependence of mass of black hole($M$) on its' charge($Q$), area($A$) and angular momentum($J$).\\
In large horizon area limit, it turns out that leading order fluctuations of charge($(\Delta Q)^{2}$) and angular momentum($(\Delta J)^{2}$) are independent of its' charge($Q$) and angular momentum($J$). This implies even a black hole with vanishingly small charge($Q$) and angular momentum($J$) can have finite fluctuations in respective quantities. This is a unique feature of our analysis and it may imply a lot in future regarding the quantizating of gravity.

\end{document}